# Derivation of Correlation Dimension from Spatial Autocorrelation Functions

Yanguang Chen

(Department of Geography, College of Urban and Environmental Sciences, Peking University, Beijing 100871, PRC. E-mail: chenyg@pku.edu.cn)

**Abstract**: Spatial autocorrelation coefficients such as Moran's index proved to be an eigenvalue of the spatial correlation matrixes. An eigenvalue represents a kind of characteristic length for quantitative analysis. However, if a spatial correlation is based on self-organized evolution, complex structure, and the distributions without characteristic scale, the eigenvalue will be ineffective. In this case, the single Moran index cannot lead to reliable statistic inferences. This paper is devoted to finding advisable approach to measure spatial autocorrelation for the scale-free processes of complex systems by means of mathematical reasoning and empirical analysis. Based on relative step function as spatial contiguity function, a series of ordered spatial autocorrelation coefficients are converted into the corresponding spatial autocorrelation functions. Then the mathematical relation between spatial correlation dimension and spatial autocorrelation functions is derived by decomposition of spatial autocorrelation functions. As results, a set of useful mathematical models are constructed for spatial analysis. Using these models, we can utilize spatial correlation dimension to make simple spatial autocorrelation analysis, and use spatial autocorrelation functions to make complex spatial autocorrelation analysis for geographical phenomena. This study reveals the inherent association of fractal patterns with spatial autocorrelation processes in nature and society. The work may inspire new ideas of spatial modeling and analysis for complex geographical systems.

# 1 Introduction

One of the keys to the method of data analysis is covariance, which reflects the joint variability of two random variables. In statistics, covariance is defined as the mean value of the product of the deviations of two random variables from their respective means. The application of covariance is extended to two directions. One is correlation coefficient, which can be treated as standardized covariance, and the other is correlation function, which can regarded as generalized covariance. A number of measures have been derived from correlation coefficient, including autocorrelation coefficient, partial correlation coefficient, part correlation coefficient, autocorrelation function, partial autocorrelation function, and spatial autocorrelation coefficient. The typical spatial autocorrelation coefficient for 2-dimensioanl space is Moran's index (Moran, 1948). An ordered set of autocorrelation coefficients can form an autocorrelation function, which is associated with a correlation function. Today, correlation function is the basis of multifractal analysis because the global fractal dimension is based on Renyi entropy and generalized correlation function (Chen, 2013; Chen and Feng, 2017; Feder, 1988; Grassberger, 1983; Grassberger, 1985; Halsey *et al*, 1986; Hentschel and Procaccia, 1983; Vicsek, 1989). In theory, the spatial analyses based on correlation coefficients and those based on correlation functions should reach the same goal by different routes, and thus can be integrated into a logical framework. However, how to establish the relationships between spatial autocorrelation coefficients and fractal dimensions is still not clear enough.

Where geographical research is concerned, spatial data analyses rely heavily on spatial correlation, including autocorrelation and cross-correlation. The precondition of using traditional statistical methods to analyze spatial data is that there is no correlation between spatial sampling points (Florax and Rey, 1995; Odland, 1988). Otherwise, the probability structure of spatial samples is not determinate, and thus the conventional statistical methods such as regression analysis and principal component analysis will be not credible. In this case, spatial autocorrelation modeling is always employed to make data analysis (Anselin, 1995; Cliff and Ord, 1973; Cliff and Ord, 1981; Goodchild, 1986; Griffith, 2003; Haggett *et al*, 1977; Lee and Li, 2017). The common spatial autocorrelation measures include Moran's index (Moran, 1948; Moran, 1950), Geary's coefficient (Geary, 1954), and Getis-Ord's index (Getis, 2009; Getis and Ord, 1992). However, in the process of spatial analysis, we encounter a paradox. This paradox may suggests the uncertainty principle of spatial correlation. If there is no spatial autocorrelation among a group of spatial elements, the spatial autocorrelation coefficient is reliable and equal to zero. On the contrary, if there is spatial autocorrelation, the values of spatial autocorrelation indicators such as Moran's index will be incredible. The calculation of the spatial correlation coefficient depends on the mean or even the standard deviation (Chen, 2013). The mean is based on the sum of observational values. Spatial autocorrelation implies that the whole is not equal to the sum of its parts, and therefore the mean and standard deviation are not affirmatory. As a result, the value of spatial autocorrelation coefficients will significantly deviate from the confidence values. One way to solve the above problem is the integration analysis of multiple correlation measures. Today, there are many measurements can be used to make spatial correlation analysis. Among various spatial correlation statistics, Moran's index and spatial correlation dimension are important ones. In order to integrate these different correlation measures, we had better reveal the logic relations between them. The aim of this paper is at deriving the inherent association of spatial autocorrelation coefficient with spatial correlation dimension. In Section 2, the concepts and models of spatial correlation functions and spatial correlation dimension are clarified, and the then spatial correlation dimension is derived from spatial autocorrelation functions based on Moran's index. In Section 3, to verify the theoretical results, the derived models are applied to the Chinese cities. In Section 4, the related questions are discussed. Finally, the discussion is concluded by summarizing the main points of this work.

# 2 Theoretical models

## 2.1 Spatial correlation dimension

Spatial correlation dimension is defined on the basis of spatial correlation function. Correlation functions can be divided into two types: correlation density function and correlation sum function (Williams, 1997). The former is based on density distribution function, and the latter is based on cumulative distribution function. In urban science, spatial correlation density function is also termed density-density correlation function, which can be expressed as follows

$$c(r)=\int_{-\infty}^{\infty}\rho(x)\rho(x+r)\mathrm{d}x\,, \tag{1}$$

where $c(r)$ refers to the density correlation, $\rho(x)$ denotes city density, $x$ is the location of a certain city (defined by the radius vector), and $r$ is the distance to $\boldsymbol{x}$ and it represents spatial displacement parameter. In terms of equation (1), if there is a city at ***x***, the probability to find another city at distance $r$ from $x$ is $c(r)$. The correlation function based on integral is useful in theoretical deduction. In application, the continuous form should be replaced by discrete form, which can be expressed as

$$c(r)=\frac{1}{S}\sum_{x}\rho(x)\rho(x+r)\,, \tag{2}$$

where $S$ denotes the area of a geographical unit occupied by a system of cities. The other symbols are the same as those in equation (1). If we can find the relationship between the correlation function $c(r)$ and the spatial displacement $r$, we can make a spatial analysis of cities. Equation (2) is the discrete expression of density-density correlation function. Through integral, it can be transformed into a correlation sum function as below (Chen, 2008b; Chen and Jiang, 2010):

$$C(r)=\frac{1}{S}\sum_{x}A(x)A(x+r)\,, \tag{3}$$

where $C(r)$ is called **correlation integral** or **correlation sum** (Williams, 1997), $A(x)$ denotes urban mass. The density correlation is a decreasing function, while the mass correlation is an increasing function. Correlation density functions are susceptible to random perturbations. In contrast, cumulative function has strong anti-noise ability, and thus can better reflect the spatial regularity.

In practice, if we use the categorical (nominal) variable to substitute the metric variable, the correlation sum function can be further simplified. Based on spatial nominal variable, equation (3) can be rewritten as

$$C(r)=\frac{N(r)}{N^2}=\frac{1}{N^2}\sum_{i=1}^{N}\sum_{j=1}^{N}H(r-d_{ij})\,, \tag{4}$$

which $r$ refers to the yardstick indicative of distance threshold, $N$ denotes the number of all cities in the study area, $N(r)$ is the number of the cities have correlation, $d_{ij}$ is the distance between city $i$ and city $j$ ($i$, $j$=1,2,3,…,$N$), and $H(\bullet)$ is the Heaviside function. The property of Heaviside function is as below

$$H(r-d_{ij})=\begin{cases}1, & \text{when}\, d_{ij}\le r;\\ 0, & \text{when}\, d_{ij}>r.\end{cases} \tag{5}$$

This implies that $r$ forms a distance yardstick by the Heaviside function. If the relationship between correlation sum and the distance threshold follow a power law such as

$$C(r)=C_1 r^{D_c}, \tag{6}$$

we will have a scale-free correlation, and $D_c$ is the correlation dimension coming between 0 and 2. In equation (6), $C_1$ refers to the proportionality coefficient. In empirical analyses, the correlation sum $C(r)$ can be replaced by correlation number $N(r)$ to determine fractal dimension. Obviously, the correlation number is

$$N(r)=\sum_{i=1}^{N}\sum_{j=1}^{N}H(r-d_{ij}). \tag{7}$$

Then equation (6) should be substituted with the following relation

$$N(r)=N^2C(r)=N_1 r^{D_c}, \tag{8}$$

where $N_1= C_1N^2$ denotes the proportionality coefficient. Replacing the correlation function $C(r)$ with the correlation number $N(r)$ has no influences on the value of the spatial correlation dimension, $D_c$. In this case, equation (8) is actually equivalent to equation (6) in geographical spatial analysis.

## 2.2 Spatial autocorrelation function based on Moran's *I*

Generalizing spatial autocorrelation coefficients yields corresponding spatial autocorrelation functions. Introduction of variable distance into spatial contiguity matrix may yield ordered sets of spatial autocorrelation coefficient (Bjørnstad and Falck, 2001; Getis and Ord, 1992; Legendre and Legendre, 1998; Odland, 1988). The autocorrelation coefficient sets can be developed into spatial autocorrelation functions. Spatial autocorrelation coefficients are determined by size measures and spatial proximity measures. A spatial proximity matrix, which is a spatial distance matrix or a spatial relation matrix, can be converted into a contiguity matrix as follows

$$V=\left[v_{ij}\right]_{N\times N}=\begin{bmatrix} v_{11} & v_{12} & \cdots & v_{1N} \\ v_{21} & v_{22} & \cdots & v_{2N} \\ \vdots & \vdots & \ddots & \vdots \\ v_{N1} & v_{N2} & \cdots & v_{NN} \end{bmatrix}. \tag{9}$$

The spatial contiguity can be defined by a step function (Lee and Li, 2017; Legendre and Legendre, 1998). There are two types of step function in geographical analysis, that is, absolute step function and relative step function. The former is based on fixed distance threshold, and the letter is based on variable distance threshold. The relative step function can be expressed as below

$$v_{ij}(r)=\begin{cases}1, & 0<d_{ij}\le r\\ 0, & d_{ij}>r\end{cases}, \tag{10}$$

where $d_{ij}$ refers to the distance between locations $i$ and $j$, $r$ denotes a variable distance threshold. The distance threshold $r$ is just the yardstick for computing the spatial correlation dimension, and it represents the displacement parameter in spatial autocorrelation functions. If $i=j$ indicates $v_{ij}(r)=0$, then it follows

$$M(r)=\left[v_{ij}(r)\right]_{N\times N}=\begin{bmatrix}0 & v_{12}(r) & \cdots & v_{1N}(r)\\ v_{21}(r) & 0 & \cdots & v_{2N}(r)\\ \vdots & \vdots & \ddots & \vdots\\ v_{N1}(r) & v_{N2}(r) & \cdots & 0\end{bmatrix}. \tag{11}$$

This is one basis for conventional spatial autocorrelation analysis. On the other, if $i=j$ suggests $v_{ij}(r)=1$, then we will have

$$M^*(r)=\left[v_{ij}(r)\right]_{N\times N}=\begin{bmatrix}1 & v_{12}(r) & \cdots & v_{1N}(r)\\ v_{21}(r) & 1 & \cdots & v_{2N}(r)\\ \vdots & \vdots & \ddots & \vdots\\ v_{N1}(r) & v_{N2}(r) & \cdots & 1\end{bmatrix}. \tag{12}$$

This will be used to make generalized spatial autocorrelation analysis. Obviously, the difference between $M^*(r)$ and $M(r)$ is a unit matrix $E$, that is

$$M^*(r)-M(r)=E. \tag{13}$$

The sum of the elements in the contiguity matrix is as follows

$$T(r)=\sum_{i=1}^{n}\sum_{j=1}^{n}v_{ij}(r)=\begin{cases}M_0(r), & v_{ii}=0\\ M_0^*(r), & v_{ii}=1\end{cases}. \tag{14}$$

Define a constant vector $e=[1, 1, \ldots, 1]^{\mathrm{T}}$, which is also termed the $n$-by-1 vector of ones (De Jong *et al*, 1984; Dray, 2011), we have

$$M_0(r)=e^{\mathrm{T}}M(r)e, \tag{15}$$

$$M_0^*(r)=e^{\mathrm{T}}M^*(r)e. \tag{16}$$

Apparently, $N=e^{\mathrm{T}}Ee$. Thus the number of non-zero elements in the matrix $M(r)$ is

$$N(r)=M_0^*(r)=M_0(r)+N=e^{\mathrm{T}}M^*(r)e. \tag{17}$$

According to equation (7), $N(r)$ is just the correlation number of cities. In order to unitize the spatial contiguity matrix, we can define

$$\frac{v_{ij}(r)}{T(r)}=\frac{v_{ij}(r)}{\sum_{i=1}^{n}\sum_{j=1}^{n}v_{ij}(r)}=\begin{cases}w_{ij}(r),\ v_{ii}(r)=0\\ w_{ij}^{*}(r),\ v_{ii}(r)=1\end{cases}. \tag{18}$$

Thus we have

$$W(r)=\frac{M(r)}{M_0(r)}=\left[w_{ij}(r)\right]_{n\times n}, \tag{19}$$

$$W^{*}(r)=\frac{M^{*}(r)}{M_0^{*}(r)}=\left[w_{ij}^{*}(r)\right]_{n\times n}. \tag{20}$$

With the preparation of the above definitions and symbolic system, we can define the spatial autocorrelation function. Based on standardized size vector $z$ and global unitized spatial weight matrix $W$, Moran's index of spatial autocorrelation can be expressed as (Chen, 2013)

$$I=z^{\mathrm{T}}Wz. \tag{21}$$

Replacing the determined unitized spatial weight matrix $W$ by the variable unitized spatial weight matrix $W(r)$ yields

$$I(r)=z^{\mathrm{T}}W(r)z, \tag{22}$$

which is a spatial autocorrelation function of displacement based on Moran's index.

The conventional spatial autocorrelation coefficient, Moran's $I$, is obtained by analogy with the temporal autocorrelation function in the theory of time series analysis. For time series analysis, if time lag is zero ($\tau$=0), the autocorrelation coefficient reflects the self-correlation of a variable at time $t$ to the variable at time $t$. In this case, the autocorrelation coefficient must be equal to 1, a known number, and thus yields no any useful information. As a result, the zero time lag is not taken into account in time series analysis. The diagonal elements of the spatial contiguity matrix correspond to the zero lag of the time series. Accordingly, the values of the diagonal elements of the spatial contiguity matrix is always set as 0. As a matter of fact, the diagonals represent the self-correlation of spatial elements in a geographical system, e.g., city A correlates with city A, city B correlates with city B. This kind of influence cannot be ignored in many cases. If we consider the self-correlation of geographical elements, Moran's index can be generalized to the following form

$$I^{*}(r)=z^{\mathrm{T}}W^{*}(r)z. \tag{23}$$

In the spatial weight matrix $W^{*}(r)$, the values of the diagonal elements are 1. In short, spatial autocorrelation differs from temporal autocorrelation, and the diagonal elements of spatial contiguity matrix can be taken into consideration in some cases.

## 2.3 Derivation of correlation dimension from spatial autocorrelation function

If a geographical process of spatial autocorrelation has characteristic scales, we will have certain values of Moran's index. At least, we can find typical value for Moran's index. In this instance, the spatial correlation function is not necessary. In fact, the spatial contiguity matrix based on variable distance was often employed to find characteristic scale for spatial correlation coefficients (Legend and Legend, 1998; Odland, 1988). On the contrary, if a geographical correlation process bear no characteristic scale, the spatial autocorrelation function suggests scaling process in the geographical pattern. Scaling is one of necessary conditions for fractal structure (Mandelbrot, 1982). Thus, maybe we can find the fractal properties in spatial autocorrelation. By means of the concepts of spatial correlation functions and spatial autocorrelation functions, the relations between Moran's index and fractal dimension can be derived. The expression of the spatial autocorrelation function based on Moran's index can be decomposed as

$$I(r)=z^{\mathrm{T}}\left(\frac{M(r)}{M_0(r)}\right)z=z^{\mathrm{T}}\left(\frac{M^*(r)}{M_0(r)}-\frac{E}{M_0(r)}\right)z=\frac{1}{M_0(r)}(z^{\mathrm{T}}M^*(r)z-N), \tag{24}$$

in which the total number of all elements in a given geographical system can be expressed as (Chen, 2013)

$$N=z^{\mathrm{T}}Ez=z^{\mathrm{T}}z. \tag{25}$$

Thus, equation (24) can be rewritten as

$$M_0(r)I(r)=z^{\mathrm{T}}M^*(r)z-N. \tag{26}$$

The two sides of equation (26) divided by the correlation number $N(r)$ at the same time yields

$$\frac{M_0(r)I(r)}{N(r)}=z^{\mathrm{T}}\frac{M^*(r)}{M_0^*(r)}z-\frac{N}{N(r)}=z^{\mathrm{T}}W^*(r)z-\frac{N}{N(r)}=I^*(r)-\frac{N}{N(r)}. \tag{27}$$

This suggests that the autocorrelation function based on the generalized Moran's index can be decomposed as follows

$$I^*(r)=z^{\mathrm{T}}W^*(r)z=\frac{N}{N(r)}+\frac{M_0(r)I(r)}{N(r)}=\frac{z^{\mathrm{T}}Ez+z^{\mathrm{T}}M(r)z}{e^{\mathrm{T}}M^*(r)e}. \tag{28}$$

From equation (27) it follows

$$I^*(r)-\frac{M_0(r)I(r)}{M_0(r)+N}=I^*(r)-\frac{I(r)}{1+N/M_0(r)}=\frac{N}{N(r)}. \tag{29}$$

Substituting equation (8) into equation (29) yields

$$f(I(r))=I^*(r)-\frac{I(r)}{1+N/M_0(r)}=\frac{N}{N(r)}=\frac{N}{N_1}r^{-D_c}, \tag{30}$$

in which $f(I(r))$ refers to the generalized correlation function based on Moran's $I$. Equation (30) gives the mathematical relationships between the spatial autocorrelation function, $I(r)$, the generalized autocorrelation function, $I^*(r)$, and the spatial correlation dimension, $D_c$. Considering equation (4), $C(r)=N(r)/N^2$, we have

$$\frac{1}{C(r)}=NI^*(r)-\frac{N}{1+N/M_0(r)}I(r)=\frac{N^2}{N_1}r^{-D_c}. \tag{31}$$

This indicates that the relationships between spatial correlation functions and spatial autocorrelation functions are as follows

$$C(r)=\frac{1}{NI^*(r)-\frac{N}{1+N/M_0(r)}I(r)}. \tag{32}$$

With the increase of $r$, $N/M_0(r)$ approaches 0. Thus we have approximate expression as below:

$$\Delta I(r)=I^*(r)-I(r)\approx\frac{N}{N(r)}=\frac{1}{NC(r)}=\frac{N}{N_1}r^{-D_c}=\frac{1}{NC_1}r^{-D_c}, \tag{33}$$

where $\Delta I(r)$ denotes the difference between $I^*(r)$ and $I(r)$. The spatial correlation function can be approximately expressed as

$$C(r)\approx\frac{1}{N\Delta I(r)}=\frac{1}{N[I^*(r)-I(r)]}. \tag{34}$$

Up to now, we have derived the exact and approximate relationships between spatial correlation dimension and spatial autocorrelation function. The spatial correlation function comprises a series of spatial autocorrelation coefficients based on Moran's index. Using observational data, we can testify the main relations derived from the theoretical principle of spatial correlation processes.

## 2.4 Model extension

The above mathematical process suggests that, based on the relative step function of distance, spatial autocorrelation coefficients can be generalized to spatial autocorrelation functions. The typical spatial autocorrelation coefficient is Moran's index. The spatial autocorrelation function on the basis of Moran's index can be expressed as equation (22). Taking into account the self-correlation of geographical elements, the standard spatial autocorrelation function can be generalized to the form of equation (23). Equations (22) and (23) proved to be associated with the reciprocal of spatial correlation functions. The spatial correlation dimension $D_c$ can be derived from the standard spatial autocorrelation function $I(r)$ and the generalized spatial autocorrelation function, $I^*(r)$. Thus, the mathematical relationships between fractal dimension, autocorrelation coefficients, and spatial correlation dimension have been brought to light. Moreover, the spatial correlation dimension can be linked to Geary's coefficient and Getis-Ord's index. The relationship between Moran's index and Geary's coefficient can be demonstrated as

$$C=\frac{n-1}{n}(e^{\mathrm{T}}Wz^2-z^{\mathrm{T}}Wz)=\frac{n-1}{n}(e^{\mathrm{T}}Wz^2-I), \tag{35}$$

where $e=[1\ 1\ \dots\ 1]^{\mathrm{T}}$, $z^2=D_{(z)}z=[z_1^2\ z_2^2\ \dots\ z_n^2]^{\mathrm{T}}$, and $D_{(z)}$ is a diagonal matrix comprising the elements of $z$. Introducing the spatial displacement parameter $r$ into equation (35) yields the autocorrelation functions based on Geary's coefficient as follows

$$C_{\mathrm{g}}(r)=\frac{n-1}{n}[e^{\mathrm{T}}W(r)z^2-I(r)], \tag{36}$$

where $C_{\mathrm{g}}(r)$ denotes Geary's function, and the right subscript g is used to differentiate Geary's function from spatial correlation function. Considering equations (8) and (17), and then rewriting equation (30) yields

$$I(r)=\frac{M_0^*(r)}{M_0(r)}(I^*(r)-\frac{N}{N_1}r^{-D_c})=\frac{1}{M_0(r)}(I^*(r)N_1r^{D_c}-N). \tag{37}$$

Substituting equation (37) into equation (36) yields

$$C_{\mathrm{g}}(r)=\frac{n-1}{n}[e^{\mathrm{T}}W(r)z^2+\frac{N}{M_0(r)}-\frac{N_1I^*(r)}{M_0(r)}r^{D_c}], \tag{38}$$

which gives the relationships between the spatial autocorrelation function based on Geary's coefficient and spatial correlation dimension $D_c$. If $n$ is large enough, then $(n-1)/n$ is close to 1 and $N/M_0(r)$ approaches 0, and equation (38) can be replaced by an approximation relation.

Further, we can derive the relationship between Getis-Ord's index and spatial correlation dimension. Substituting the standardized size vector, $z$, in equation (21) with the unitized size vector, $u$, we can transform the formula of the spatial autocorrelation function based on Moran's index into that based on Getis-Ord's index as follows

$$G(r) = u^{\mathrm{T}} W(r) u = \frac{1}{M_0(r)} (u^{\mathrm{T}} M(r) u - u^{\mathrm{T}} u) . \tag{39}$$

Then, replacing $W(r)$ with $W^*(r)$, we can generalized standard spatial autocorrelation function to the following form

$$G^*(r) = u^{\mathrm{T}} W^*(r) u = \frac{u^{\mathrm{T}} u}{N(r)} + \frac{M_0(r) G(r)}{N(r)} = \frac{u^{\mathrm{T}} E u + u^{\mathrm{T}} M(r) u}{e^{\mathrm{T}} M^*(r) e} , \tag{40}$$

in which $u^{\mathrm{T}}u$ is a constant. Similar to the process of derivation of the relationships between Moran's index and spatial correlation dimension, a relation between Getis-Ord's index $G$ and fractal dimension $D_c$ can be derived as

$$f(G(r)) = G^*(r) - \frac{G(r)}{1 + N / M_0(r)} = \frac{u^{\mathrm{T}} u}{N(r)} = \frac{u^{\mathrm{T}} u}{N^2 C(r)} = \frac{u^{\mathrm{T}} u}{N_1} r^{-D_c} , \tag{41}$$

where $N(r)=N^2C(r)$ and $f(G(r))$ denotes the generalized correlation function based on Getis-Ord's $G$. Accordingly, an approximate relation is as below:

$$G^*(r) - G(r) \approx \frac{u^{\mathrm{T}} u}{N_1} r^{-D_c} . \tag{42}$$

So far, the common spatial autocorrelation coefficients, including Moran's index, Geary's coefficient, and Getis-Ord's index, have been generalized to spatial autocorrelation functions. All these spatial autocorrelation functions have been associated with spatial correlation dimension. Thus, Based on the ideas from fractals, three types of spatial autocorrelation measurements have been integrated into the same logic framework of spatial analysis (Table 1).

**Table 1 The main mathematical relations between spatial correlation dimension and spatial autocorrelation statistics**

| Statistic | Relation | Formula |
|---|---|---|
| **Moran's *I*** | Exact relation | $I^*(r) - \frac{I(r)}{1 + N / M_0(r)} = \frac{N}{N_1} r^{-D_c}$ |
| | Approximation relation | $I^*(r) - I(r) \approx \frac{N}{N_1} r^{-D_c}$ |
| **Getis-Ord's *G*** | Exact relation | $G^*(r)-\frac{G(r)}{1+N/M_0(r)}=\frac{u^{\mathrm{T}}u}{N_1}r^{-D_c}$ |
| | Approximation relation | $G^*(r)-G(r)\approx\frac{u^{\mathrm{T}}u}{N_1}r^{-D_c}$ |
| **Geary's *C*** | Exact relation | $C_g(r)=\frac{n-1}{n}[e^{\mathrm{T}}W(r)z^2+\frac{N}{M_0(r)}-\frac{N_1I^*(r)}{M_0(r)}r^{D_c}]$ |
| | Approximation relation | $C_g(r)\approx e^{\mathrm{T}}W(r)z^2-\frac{N_1I^*(r)}{M_0(r)}r^{D_c}$ |

The derivation results suggest that the spatial correlation dimension reflect both the spatial autocorrelation and spatial interaction. Moran's index is a spatial correlation coefficient, Geary's coefficient is a spatial Durbin-Watson statistic, while Getis-Ord's index proved to be equivalent to the potential formula under certain conditions. Moran's index and Geary's coefficient reflect the extent and property of spatial autocorrelation, while Getis-Ord's index reflect both the spatial autocorrelation and spatial interaction. All these spatial statistics are associated with the spatial correlation dimension. In this sense, the spatial correlation dimension contain two aspects of geographical spatial information: spatial autocorrelation and spatial interaction.

# 3 Empirical analysis

## 3.1 Datasets and methods

The network of Chinese cities can be employed to verify the models derived in last section. For comparability and simplifying the analytical processes, only municipalities directly under the Central Government of China and provincial capitals are taken into account in this case. There are 31 provinces, municipalities, and autonomous regions in Chinese mainland. So, this network includes 31 large cities. Basic data include urban population and railway mileage. Urban population represents city size measure, and the spatial contiguity matrix is generated by railway distances. Population data came from the fifth (2000) and sixth (2010) censuses, and railway mileage came from China's traffic mileage map. However, two cities, Lhasa and Haikou, were not connected to the network by railway for a long time. Therefore, only 29 cities compose the spatial sample ($N$=29).

The analytical procedure can be outlined according to the theoretical derivation process. The computational steps are as follows. **Step 1: define the yardsticks of spatial correlation**. The yardstick is a variable of distance threshold, which is designed in light of the railway mileage matrix. Its function bears analogy with time lag parameter in time series analysis. If the zero elements on the diagonal are overlooked, the minimum traffic mileage is 137 kilometer and the maximum traffic mileage is 5062 kilometer. So the yardstick length can be taken as $r$=150, 250, 350, …, 5150. **Step 2: calculate spatial correlation function**. Using Heaviside function, equation (5), we can obtain spatial correlation number $N(r)$, and spatial correlation function, $C(r)$. Based on scaling range, the correlation dimension can be evaluated by the power law relation between the yardstick length $r$ and spatial correlation number $N(r)$ or spatial correlation function $C(r)$. **Step 3: compute spatial autocorrelation measurements based on variable yardstick**. The spatial autocorrelation measures include Moran's index, Geary's coefficient, and Getis-Ord's index. This work is mainly based on Moran index, supplemented by Geary coefficient and Getis-Ord's index. **Step 4: verify the relationship between spatial autocorrelation measures and fractal dimension**. Using equations (30) and (33), we can confirm the relationships between Moran's index and spatial correlation dimension. In theory, this positive study can be generalized to the relationships between fractal dimension and Geary's coefficient and Getis-Ord's index.

Analytical process and results depend heavily on the definition and structure of spatial weight matrix. Where structure is concerned, two aspects of factors significantly influence analytical ways. One is diagonal elements, and the other is sum of spatial contiguity matrix. For fractal analysis, the diagonal elements should be taken into account, while for conventional spatial autocorrelation analysis, the diagonal elements should be removed. For generalized spatial autocorrelation analysis, the diagonal elements can be taken into consideration, while for special fractal analysis, the diagonal element can be deleted. On the other hand, for practical spatial autocorrelation function, the sum of spatial contiguity matrix should be fixed to the original sum value. However, for theoretical spatial autocorrelation function, the sum varies with the yardstick length. Different sums of spatial contiguity matrix plus different diagonal elements lead to four approaches to spatial correlation dimension and autocorrelation analyses (Table 2).

**Table 2 Four types of calculation approaches to spatial autocorrelation measurements**

| | **Variable sum of distance matrix [V]** | **Fixed sum of distance matrix [F]** |
|---|---|---|
| **All elements (including diagonal elements) [D]** | [D+V] Generalized Moran's function, $I^*(r)$ | [D+F] Generalized Moran's function, $I_f^*(r)$ |
| **Partial elements (excluding diagonal elements) [N]** | [N+V] Conventional Moran's function, $I(r)$ | [N+F] Conventional Moran's function, $I_f(r)$ |
| **Application direction** | Theoretical study and fractal analysis | Practical study and spatial autocorrelation analysis |

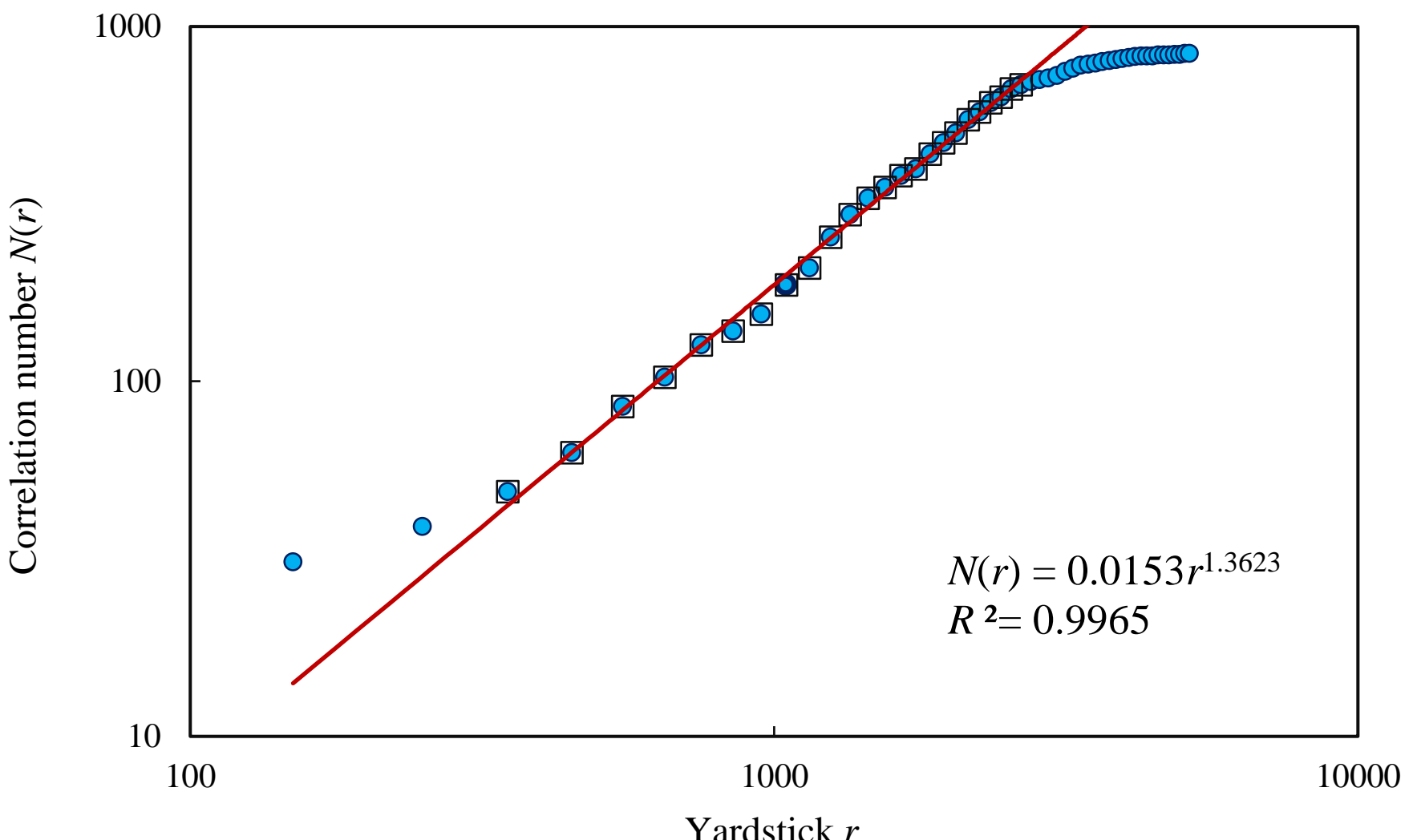

**Figure 1 The scaling relation for spatial correlation dimension of Chinese provincial capital cities based on railway distance**

**Note:** The solid dots represent the total number of spatial correlations, and the hollow blocks represent the points within the scaling range (250km<scaling range<2750 km). The latter is a subset of the former.

## 3.2 Computed results and analysis based on Moran's *I*

Using the data and methods, we can testify the models proposed above. In fractal analysis, scaling relationships take on two forms: one is global scaling, and the other is local scaling. The global scaling relations imply that all data points follow power law and form a straight line on the double logarithmic plot. In contrast, the local scaling relations indicate that only part data points follow power law and form a local straight line segment on the log-log plot. In theory, all the scaling relations are global power law relations, but empirically, almost all scaling relationships are local power law relations. In many cases, if the linear scale for measurement is too large or too small, the power law relations break (Bak, 1996). The local straight line segment represents the scaling range for fractal analysis. Partial calculation results are tabulated as below (Table 3). If the yardstick length is less than 300 milometers or greater than 2700 milometers, the power law relations break. The scaling range varies from 350 milometers to 2650 milometers (Figure 1). The relation between yardstick length $r$ and the correlation number $N(r)$ follows the power law, and the mathematical model is as follows

$$\hat{N}(r) = 0.0153r^{1.3623} . \tag{43}$$

The goodness of fit is about $R^2$=0.9965, and the spatial correlation dimension is about $D_c$=1.3623. The symbol "^" denotes that the result is estimated value.

**Table 3 Datasets for spatial correlation dimension and spatial autocorrelation analysis (Partial results)**

| **Scale** | **Number** | | **2000 (Fifth census data)** | | | | **2010 (Sixth census data)** | | | |
|---|---|---|---|---|---|---|---|---|---|---|
| ***r*** | $N(r)$ | $N^*(r)$ | Moran $I^*$ | Moran $I$ | $\Delta I$ | $1/NC(r)$ | Moran $I^*$ | Moran $I$ | $\Delta I$ | $1/NC(r)$ |
| **150** | 31 | 2 | 1.0411 | 1.6363 | -0.5953 | 0.9355 | 1.1172 | 2.8164 | -1.6992 | 0.9355 |
| **250** | 39 | 10 | 0.8015 | 0.2257 | 0.5758 | 0.7436 | 0.9139 | 0.6643 | 0.2496 | 0.7436 |
| ***350*** | 49 | 20 | 0.5907 | -0.0028 | 0.5935 | 0.5918 | 0.6931 | 0.2481 | 0.4450 | 0.5918 |
| ***450*** | 63 | 34 | 0.4130 | -0.0877 | 0.5007 | 0.4603 | 0.5008 | 0.0749 | 0.4258 | 0.4603 |
| ***550*** | 85 | 56 | 0.2876 | -0.0813 | 0.3689 | 0.3412 | 0.3303 | -0.0164 | 0.3468 | 0.3412 |
| ***650*** | 103 | 74 | 0.2158 | -0.0915 | 0.3073 | 0.2816 | 0.2670 | -0.0203 | 0.2892 | 0.2816 |
| ***750*** | 127 | 98 | 0.1681 | -0.0780 | 0.2462 | 0.2283 | 0.1948 | -0.0435 | 0.2383 | 0.2283 |
| ***850*** | 139 | 110 | 0.1065 | -0.1291 | 0.2356 | 0.2086 | 0.1215 | -0.1101 | 0.2316 | 0.2086 |
| ***950*** | 155 | 126 | 0.1080 | -0.0972 | 0.2053 | 0.1871 | 0.1250 | -0.0764 | 0.2014 | 0.1871 |
| ***1050*** | 187 | 158 | 0.0489 | -0.1257 | 0.1746 | 0.1551 | 0.0543 | -0.1193 | 0.1736 | 0.1551 |
| ***1150*** | 209 | 180 | 0.0478 | -0.1056 | 0.1534 | 0.1388 | 0.0471 | -0.1064 | 0.1535 | 0.1388 |
| ***1250*** | 255 | 226 | 0.0668 | -0.0529 | 0.1197 | 0.1137 | 0.0471 | -0.0752 | 0.1223 | 0.1137 |
| ***1350*** | 295 | 266 | 0.0357 | -0.0695 | 0.1051 | 0.0983 | 0.0314 | -0.0742 | 0.1056 | 0.0983 |
| ***1450*** | 329 | 300 | 0.0312 | -0.0624 | 0.0936 | 0.0881 | 0.0199 | -0.0748 | 0.0947 | 0.0881 |
| ***1550*** | 353 | 324 | 0.0717 | -0.0113 | 0.0831 | 0.0822 | 0.0643 | -0.0194 | 0.0837 | 0.0822 |
| ***1650*** | 381 | 352 | 0.0491 | -0.0293 | 0.0783 | 0.0761 | 0.0471 | -0.0314 | 0.0785 | 0.0761 |
| ***1750*** | 397 | 368 | 0.0372 | -0.0387 | 0.0759 | 0.0730 | 0.0359 | -0.0400 | 0.0760 | 0.0730 |
| ***1850*** | 437 | 408 | 0.0491 | -0.0185 | 0.0676 | 0.0664 | 0.0431 | -0.0250 | 0.0684 | 0.0664 |
| ***1950*** | 471 | 442 | 0.0348 | -0.0285 | 0.0633 | 0.0616 | 0.0331 | -0.0303 | 0.0634 | 0.0616 |
| ***2050*** | 501 | 472 | 0.0408 | -0.0182 | 0.0589 | 0.0579 | 0.0376 | -0.0215 | 0.0591 | 0.0579 |
| ***2150*** | 547 | 518 | 0.0179 | -0.0371 | 0.0550 | 0.0530 | 0.0151 | -0.0401 | 0.0551 | 0.0530 |
| ***2250*** | 575 | 546 | 0.0043 | -0.0486 | 0.0529 | 0.0504 | 0.0005 | -0.0526 | 0.0531 | 0.0504 |
| ***2350*** | 611 | 582 | 0.0217 | -0.0271 | 0.0487 | 0.0475 | 0.0176 | -0.0313 | 0.0490 | 0.0475 |

| ***2450*** | 633 | 604 | 0.0045 | -0.0433 | 0.0478 | 0.0458 | 0.0042 | -0.0436 | 0.0478 | 0.0458 |
|---|---|---|---|---|---|---|---|---|---|---|
| ***2550*** | 667 | 638 | 0.0175 | -0.0271 | 0.0447 | 0.0435 | 0.0171 | -0.0276 | 0.0447 | 0.0435 |
| ***2650*** | 685 | 656 | 0.0095 | -0.0343 | 0.0438 | 0.0423 | 0.0093 | -0.0345 | 0.0438 | 0.0423 |
| **2750** | 699 | 670 | 0.0047 | -0.0384 | 0.0431 | 0.0415 | 0.0030 | -0.0401 | 0.0432 | 0.0415 |
| **2850** | 709 | 680 | 0.0022 | -0.0403 | 0.0426 | 0.0409 | 0.0007 | -0.0420 | 0.0426 | 0.0409 |
| **2950** | 717 | 688 | 0.0026 | -0.0394 | 0.0420 | 0.0404 | 0.0019 | -0.0402 | 0.0421 | 0.0404 |
| **3050** | 729 | 700 | -0.0053 | -0.0470 | 0.0416 | 0.0398 | -0.0053 | -0.0470 | 0.0416 | 0.0398 |

**Note:** (1) Only partial results are displayed in this table. More results are attached in the Supporting Information files. (2) Moran's index comes between -1 and 1, otherwise the results are outliers. Corresponding to the yardstick length $r$=150, several Moran's index values are abnormal and can be treated as outliers.

A problem is how to determine the scaling range objectively for the fractal dimension estimation. This problem can be solved by the residuals sequence of global double logarithmic regression model and the goodness of fit of local double logarithmic linear regression model. The process can be illustrated as below: (1) Intuitive judgment by means of the plot of residuals based on global regression. The concept of scaling is ignored for the time being, and all the observed data are used to make double logarithmic linear regression analysis. The independent variable is ln$r$, and corresponding dependent variable is ln$N$($r$). As a result, the residuals sequence fall into three segments, and the middle segment indicates the scaling range (Figure 2). The lower limit is about 350 km, and the upper limit may be 2650 km. (2) Further judgment by the curve of goodness of fit based on the local regression. The lower limit (350km) is relatively clear, but the upper limit (2650 km) is not very certain. Thus, the coefficient of determination can be utilized to confirm the upper limit. Suppose the scaling range comes between 350 km and $d$ km, where d denotes the upper limit. The value of $d$ is taken as 550, 650, 750, …, 5150 km in turn. Changing the $d$ value yields different values of determination coefficient, i.e., $R^2$. When $d$=2550 km, we have $R^2$=0.99631; When $d$=2650 km, we have $R^2$=0.99647; When $d$=2750 km, we have $R^2$=0.99646; When $d$=2850 km, we have $R^2$=0.99621…. All in all, when $d$=2650 km, the goodness of fit, $R^2$, reached the peak of 0.99647 (Figure 3). Of course, if the upper limit of the scaling range is 2750 km, the goodness of fit is $R^2$=0.99646, and the fractal dimension is about $D_c$=1.3571. Where the scaling range limit is concerned, there is no significant difference between 2650 and 2750 km. An inference is that the lower limit of the scaling range is greater than 250 km and the upper limit is less than 2750 km. Maybe the interval ranges from 300km to 2700km. Since the numerical value of the distance yardstick is discrete, it is not necessary and possible to give an accurate scaling interval.

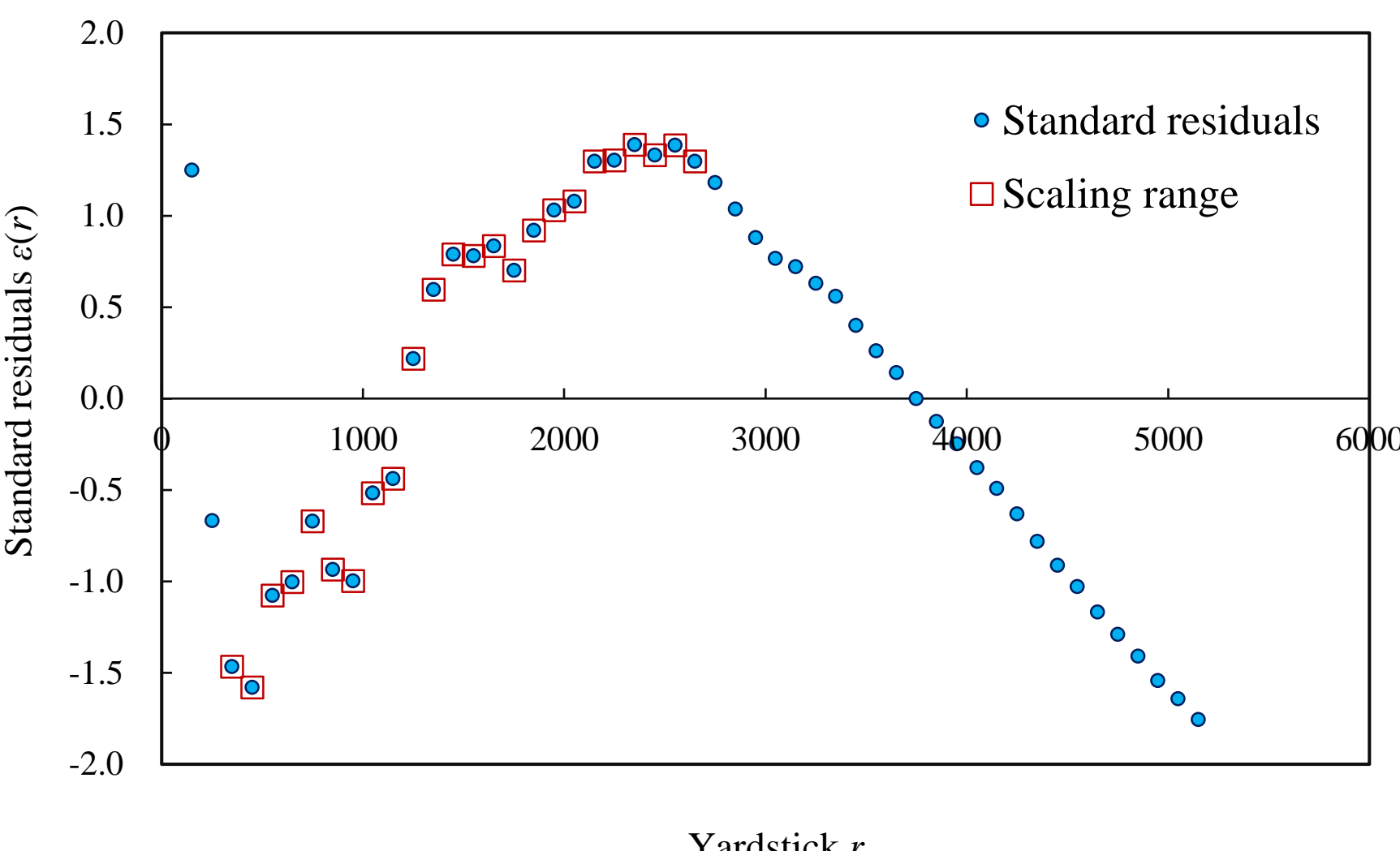

**Figure 2 The standard residuals sequences based on the double logarithmic linear regression of all the observational data of spatial correlation numbers of Chinese provincial capital cities**

**Note:** The data points of residuals can be divided into three parts. The first two points are outliers, the last part are also of exception. The second part represents the scaling range coming between 300 km and 2700 km.

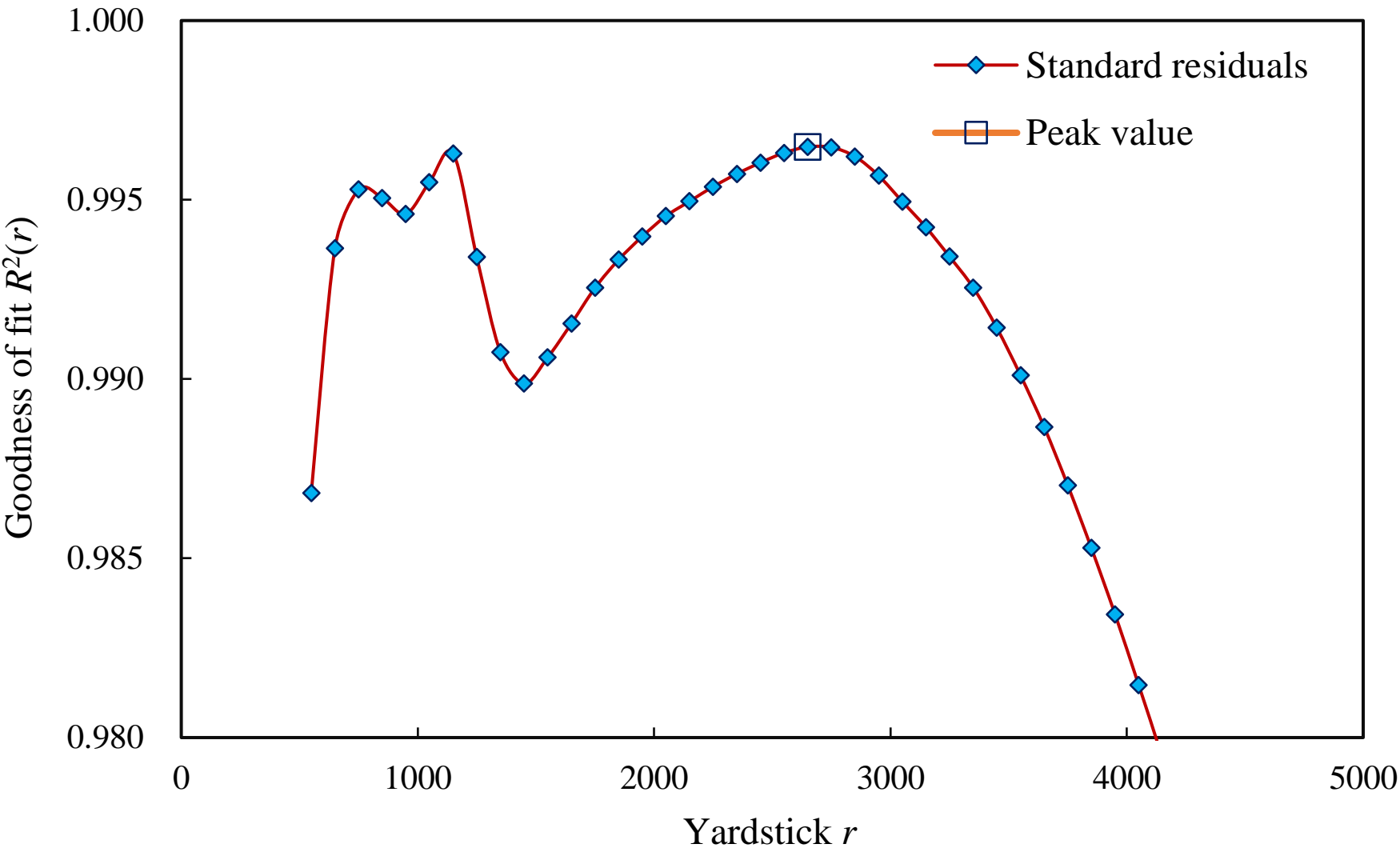

**Figure 3 The goodness of fit for the double logarithmic linear regression of partial observational data of spatial correlation numbers of Chinese provincial capital cities**

**Note:** The starting point of the scaling range is 350 km, and the terminal point is set as 550 km, 650 km, 750 km, …, 2550 km, 2650 km, 2750 km, …, and 5150 km in turn. For fewer observations, the results are unstable. When the scaling range comes between 350 km and 2650 km, the goodness of fit reached the peak of 0.99647.

The spatial correlation dimension has been theoretically associated with spatial autocorrelation functions based on conventional Moran's indexes and generalized Moran's indexes. This relation can be verified by equation (30) or equation (31). For the dataset in 2000, the mathematical model

is as below:

$$\hat{f}(I(r)) = \frac{1}{N\hat{C}(r)} = \hat{I}^*(r) - \frac{1}{1+N/M_0(r)}\hat{I}(r) = 1893.8457r^{-1.3623}. \tag{44}$$

The coefficient of determination is about $R^2$=0.9965, and the spatial correlation dimension is around $D_c$=1.3623. The fractal parameter is the same as that based on equation (43). Where spatial correlation function is concerned, this is the dimension estimation value based on an exact relation. Then, the 2010 urban census data is used to replace the 2000 urban census data, and the calculation results remain unchanged (Figure 4). The reason is that the spatial weight matrix has not changed. This suggests that the spatial scaling exponent of equation (30) or equation (31) depend on spatial contiguity matrix rather than urban population sizes. Spatial correlation dimension is only determined by spatial patterns.

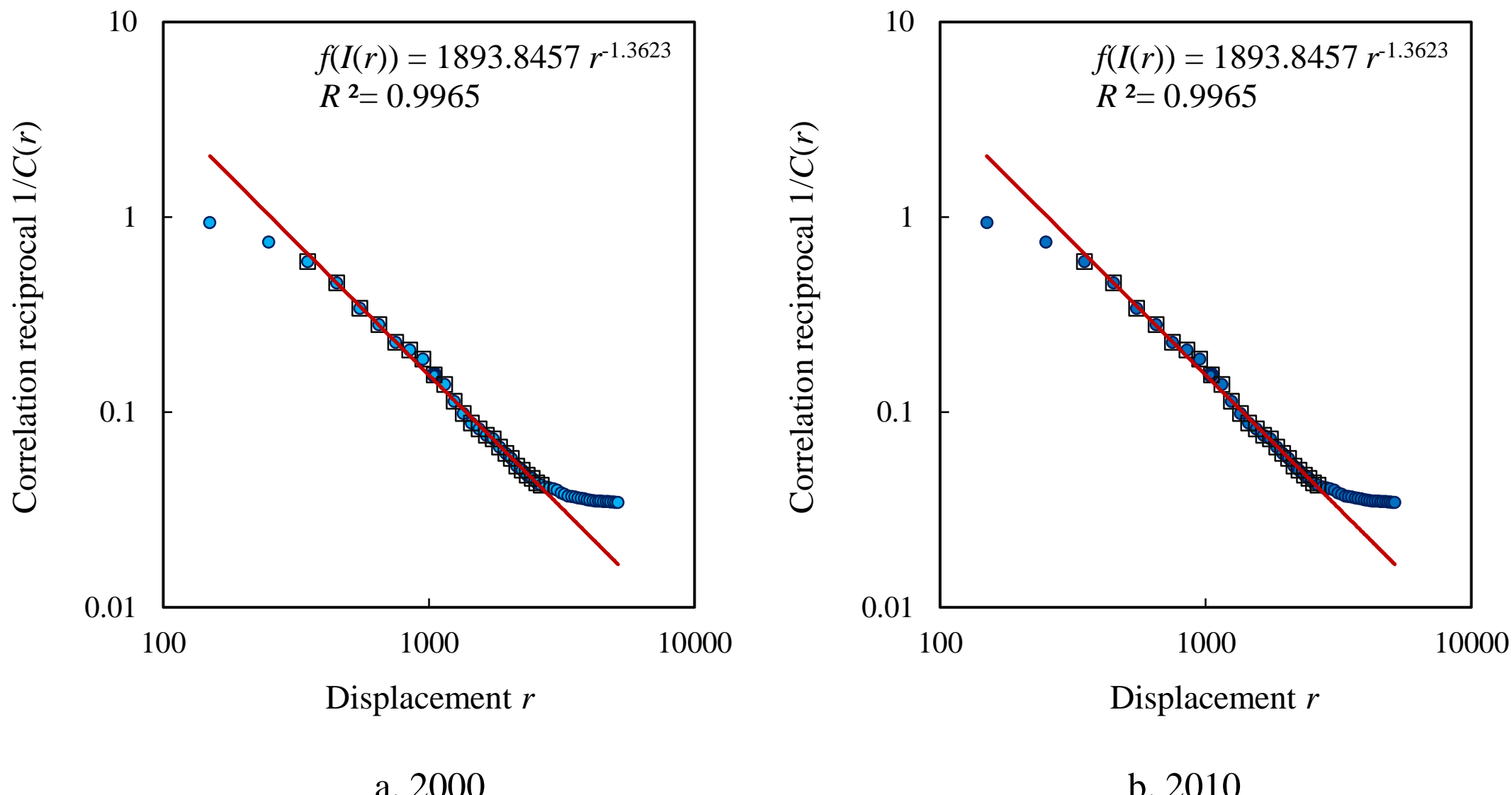

**Figure 4 The scaling relations for the reciprocal of spatial correlation function based on Moran's index**

**Note:** The solid dots represent the total number of spatial autocorrelation functions, and the hollow blocks represent the points within the scaling range. The scaling range corresponds to that in Figure 1.

If the spatial correlation number is significantly greater than the city number, the exact relation between Moran's function and yardstick length can be replaced by an approximate relation. Through equation (33), we can verify this approximate scaling relation (Figure 5). For 2000 dataset, the model based on the least square calculation is

$$\Delta\hat{I}(r) = 2423.6543r^{-1.3892}. \quad (45)$$

The goodness of fit is about $R^2$=0.9919, and the spatial correlation dimension is estimated as about $D_c$=1.3892. For 2010 data, the model is

$$\Delta\hat{I}(r) = 1229.1265r^{-1.2979}. \quad (46)$$

The goodness of fit is about $R^2$=0.9812, and the spatial correlation dimension is about $D_c$=1.2979. The goodness of fit decrease, and the fractal dimension estimation results departed from the expected value. In this case, both urban population sizes and spatial contiguity matrix influence the parameter estimation values.

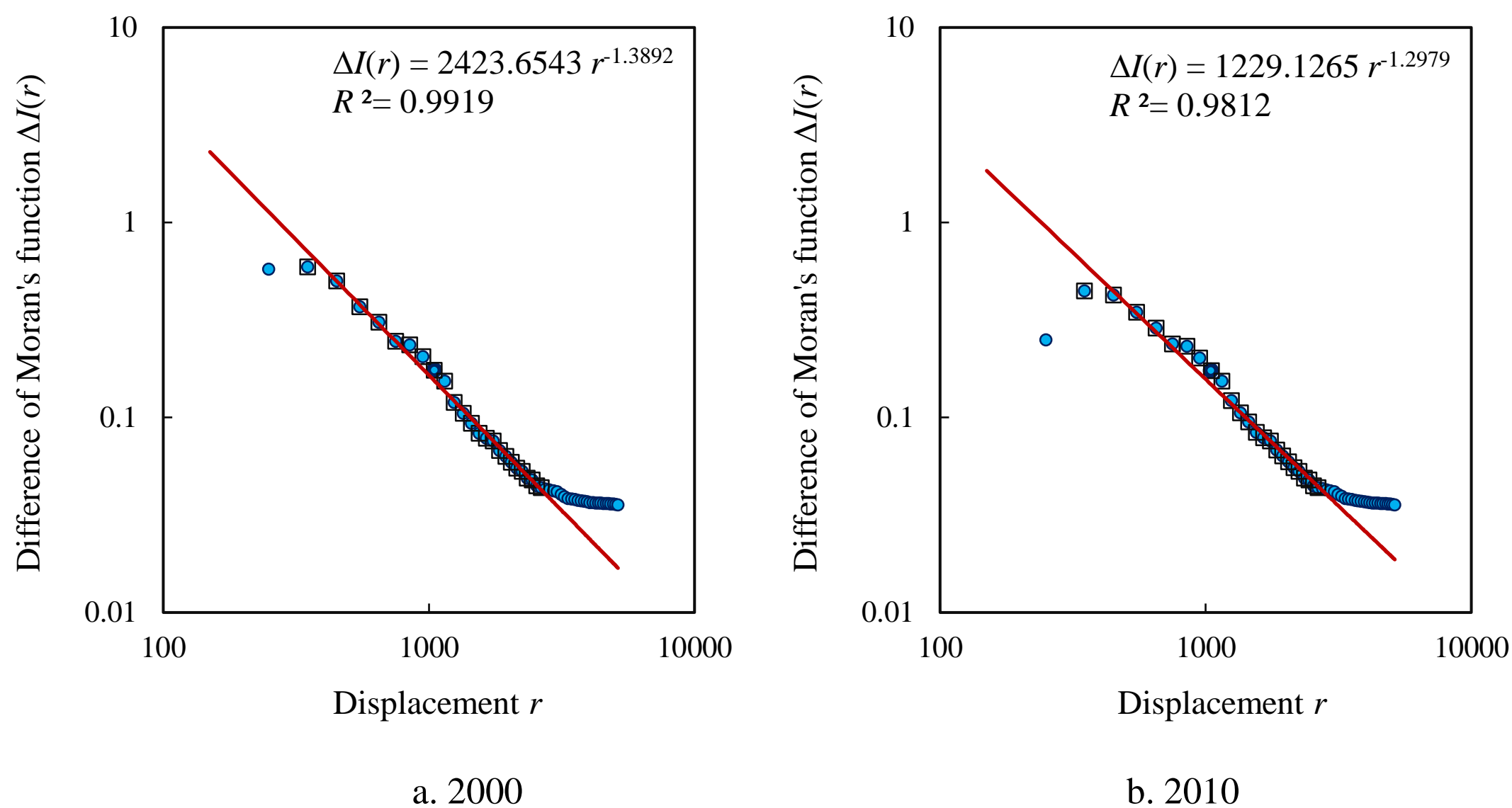

**Figure 5 The scaling relations for the difference between two types of Moran's index**

**Note:** The solid dots represent the total number of difference of Moran's functions, and the hollow blocks represent the points within the scaling range. The scaling range is consistent with those in Figures 1 and 4.

This study is devoted to exploring the theoretical relationships between spatial autocorrelation and spatial dimension. The aim at reveal the scaling in the spatial processes. The positive analysis of spatial autocorrelation and fractal dimension of urban systems is not the main task of this work. Based on the above calculation results, the inferences can be made as follows. First, spatial correlation dimension depends on spatial contiguity matrix. It is independent of size measures. Even if the city sizes changes, but the spatial distances between cities does not change, then the spatial correlation dimension remains unchanged. In this case, the relationships between Moran's function and spatial correlation dimension do not change. Second, the difference between common Moran's function and generalized Moran's function relies on both spatial contiguity matrix and size measures. If the number of cities in a region is large enough, the difference between the two Moran functions can be used to take place of the reciprocal of the correlation function. This relationships between the difference values and yardstick lengths follow power law and give spatial correlation dimension approximately. In this instance, the spatial correlation dimension value is sensitive to the city sizes. A conclusion can be drawn that theoretical spatial correlation dimension depends on the patterns of spatial distribution rather size distribution. However, if we estimate the correlation dimension using the approximate formula, the result can be impacted by the size measure.

### 3.3 Positive analyses based on Geary's *C* and Getis-Ord's *G*

The spatial autocorrelation function based on Moran's index represents a basic model of advanced spatial analysis. The auxiliary models include the spatial autocorrelation function based on Geary's coefficient and Getis-Ord's index. It is easy to calculate the spatial autocorrelation functions based on Geary's coefficient $C_g(r)$ and Getis-Ord's index $G(r)$, and the results correspond to Moran's function $I(r)$ (Table 4). In terms of equations (35) and (36), there is a strict mathematical transformation and numerical relationship between Moran's index and Geary's coefficient. In this case, it is unnecessary to testify the association of spatial correlation dimension with the spatial autocorrelation function based on Geary's *C*. However, it is helpful for understanding spatial structure of urban systems to reveal a hidden scaling relation between the spatial autocorrelation function based on Geary's coefficient and spatial displacement. Define a difference of Geary's function as follows

$$\Delta C^{*}(r)=C_{g}(r)-C_{g}^{*}(r), \qquad (47)$$

where Δ denotes difference value, $C_g(r)$ is the spatial autocorrelation function based on Geary's *C* and spatial weight matrix with zero diagonal, $C_g^*(r)$ is the spatial autocorrelation function based on Geary's *C* and spatial weight matrix with nonzero diagonal. Using the data displayed in Table 4, we can demonstrate the following power law relation

$$\Delta C^{*}(r)=Kr^{-\alpha}, \qquad (48)$$

where *K* is proportionality coefficient, and *α* is the scaling exponent. This power law relation is valid within certain scaling range (Figure 6). Based on the observational data in 2000, the model is

as below:

$$\Delta \hat{C}^{*}(r)=6698.5762r^{-1.5216}\,. \tag{49}$$

The goodness of fit is about $R^2$=0.9930, and the scaling exponent is about $\alpha$=1.5216. Based on the observational data in 2010, the model is as follows

$$\Delta \hat{C}^{*}(r)=6034.6290r^{-1.5047}\,. \tag{50}$$

The goodness of fit is about $R^2$=0.9942, and the scaling exponent is about $\alpha$=1.5047. The scaling exponent values depend on spatial distance matrix and size vector of Chinese cities. This suggests that there is no characteristic scale for the spatial autocorrelation of the system of cities in Chinese mainland.

**Table 4 Datasets for spatial autocorrelation functions based on Geary's coefficient and Getis-Ord's index (Partial results)**

| **Scale** | **2000 (Fifth census data)** | | | | **2010 (Sixth census data)** | | | |
|---|---|---|---|---|---|---|---|---|
| | D+V | | N+V | | D+V | | N+V | |
| ***r*** | Geary $C_g^*(r)$ | Getis $G^*(r)$ | Geary $C_g(r)$ | Getis $G(r)$ | Geary $C_g^*(r)$ | Getis $G^*(r)$ | Geary $C_g(r)$ | Getis $G(r)$ |
| **150** | 0.0770 | 0.0021 | 1.1934 | 0.0052 | 0.0931 | 0.0023 | 1.4432 | 0.0068 |
| **250** | 0.4366 | 0.0019 | 1.7027 | 0.0020 | 0.3687 | 0.0021 | 1.4379 | 0.0024 |
| ***350*** | 0.7660 | 0.0019 | 1.8767 | 0.0019 | 0.7144 | 0.0020 | 1.7502 | 0.0021 |
| ***450*** | 0.7343 | 0.0017 | 1.3607 | 0.0015 | 0.6769 | 0.0018 | 1.2542 | 0.0016 |
| ***550*** | 0.7619 | 0.0016 | 1.1565 | 0.0014 | 0.7835 | 0.0016 | 1.1892 | 0.0015 |
| ***650*** | 0.8146 | 0.0014 | 1.1338 | 0.0012 | 0.8068 | 0.0015 | 1.1230 | 0.0013 |
| ***750*** | 0.8517 | 0.0014 | 1.1038 | 0.0012 | 0.9123 | 0.0015 | 1.1822 | 0.0013 |
| ***850*** | 0.9366 | 0.0014 | 1.1835 | 0.0012 | 0.9996 | 0.0014 | 1.2631 | 0.0013 |
| ***950*** | 0.8701 | 0.0013 | 1.0703 | 0.0011 | 0.9148 | 0.0013 | 1.1254 | 0.0012 |
| ***1050*** | 0.9711 | 0.0013 | 1.1493 | 0.0012 | 1.0103 | 0.0013 | 1.1957 | 0.0012 |
| ***1150*** | 0.9265 | 0.0013 | 1.0757 | 0.0012 | 0.9705 | 0.0013 | 1.1268 | 0.0012 |
| ***1250*** | 0.9994 | 0.0014 | 1.1276 | 0.0013 | 1.0329 | 0.0014 | 1.1654 | 0.0013 |
| ***1350*** | 1.0589 | 0.0014 | 1.1743 | 0.0013 | 1.0789 | 0.0014 | 1.1965 | 0.0013 |
| ***1450*** | 1.0060 | 0.0013 | 1.1032 | 0.0013 | 1.0407 | 0.0013 | 1.1413 | 0.0013 |
| ***1550*** | 1.0299 | 0.0014 | 1.1221 | 0.0014 | 1.0574 | 0.0014 | 1.1520 | 0.0013 |
| ***1650*** | 1.0240 | 0.0014 | 1.1084 | 0.0013 | 1.0531 | 0.0014 | 1.1398 | 0.0013 |
| ***1750*** | 1.0118 | 0.0014 | 1.0916 | 0.0013 | 1.0367 | 0.0014 | 1.1184 | 0.0013 |
| ***1850*** | 0.9820 | 0.0013 | 1.0518 | 0.0013 | 1.0078 | 0.0013 | 1.0794 | 0.0013 |
| ***1950*** | 0.9536 | 0.0013 | 1.0162 | 0.0013 | 0.9684 | 0.0013 | 1.0319 | 0.0012 |
| ***2050*** | 0.9304 | 0.0013 | 0.9876 | 0.0013 | 0.9429 | 0.0013 | 1.0009 | 0.0013 |
| ***2150*** | 1.0066 | 0.0013 | 1.0630 | 0.0013 | 1.0163 | 0.0013 | 1.0732 | 0.0013 |
| ***2250*** | 1.0119 | 0.0013 | 1.0657 | 0.0013 | 1.0212 | 0.0013 | 1.0755 | 0.0013 |
| ***2350*** | 0.9789 | 0.0013 | 1.0277 | 0.0012 | 0.9937 | 0.0013 | 1.0432 | 0.0012 |
| ***2450*** | 1.0503 | 0.0013 | 1.1007 | 0.0013 | 1.0518 | 0.0013 | 1.1023 | 0.0013 |
| ***2550*** | 1.0254 | 0.0013 | 1.0720 | 0.0013 | 1.0275 | 0.0013 | 1.0742 | 0.0013 |
| ***2650*** | 1.0372 | 0.0013 | 1.0831 | 0.0012 | 1.0403 | 0.0013 | 1.0862 | 0.0013 |
| **2750** | 1.0434 | 0.0013 | 1.0886 | 0.0012 | 1.0477 | 0.0013 | 1.0930 | 0.0012 |
| **2850** | 1.0370 | 0.0013 | 1.0813 | 0.0012 | 1.0408 | 0.0013 | 1.0852 | 0.0012 |
| **2950** | 1.0271 | 0.0013 | 1.0704 | 0.0012 | 1.0304 | 0.0013 | 1.0738 | 0.0012 |
| **3050** | 1.0329 | 0.0013 | 1.0757 | 0.0012 | 1.0326 | 0.0013 | 1.0753 | 0.0012 |

**Note:** (1) The yardstick length $r$ represents measurement scales and displacement parameter of spatial correlation. (2) Difference scales $r$ lead to different Geary's coefficients $C$ and Getis-Ord's index $G$, which form Geary's function $C_g(r)$ and Getis-Ord's function $G(r)$. (3) D implies that diagonal elements are taken into account, N means that diagonal elements are removed, and V denotes variable mean values of spatial contiguity matrix elements.

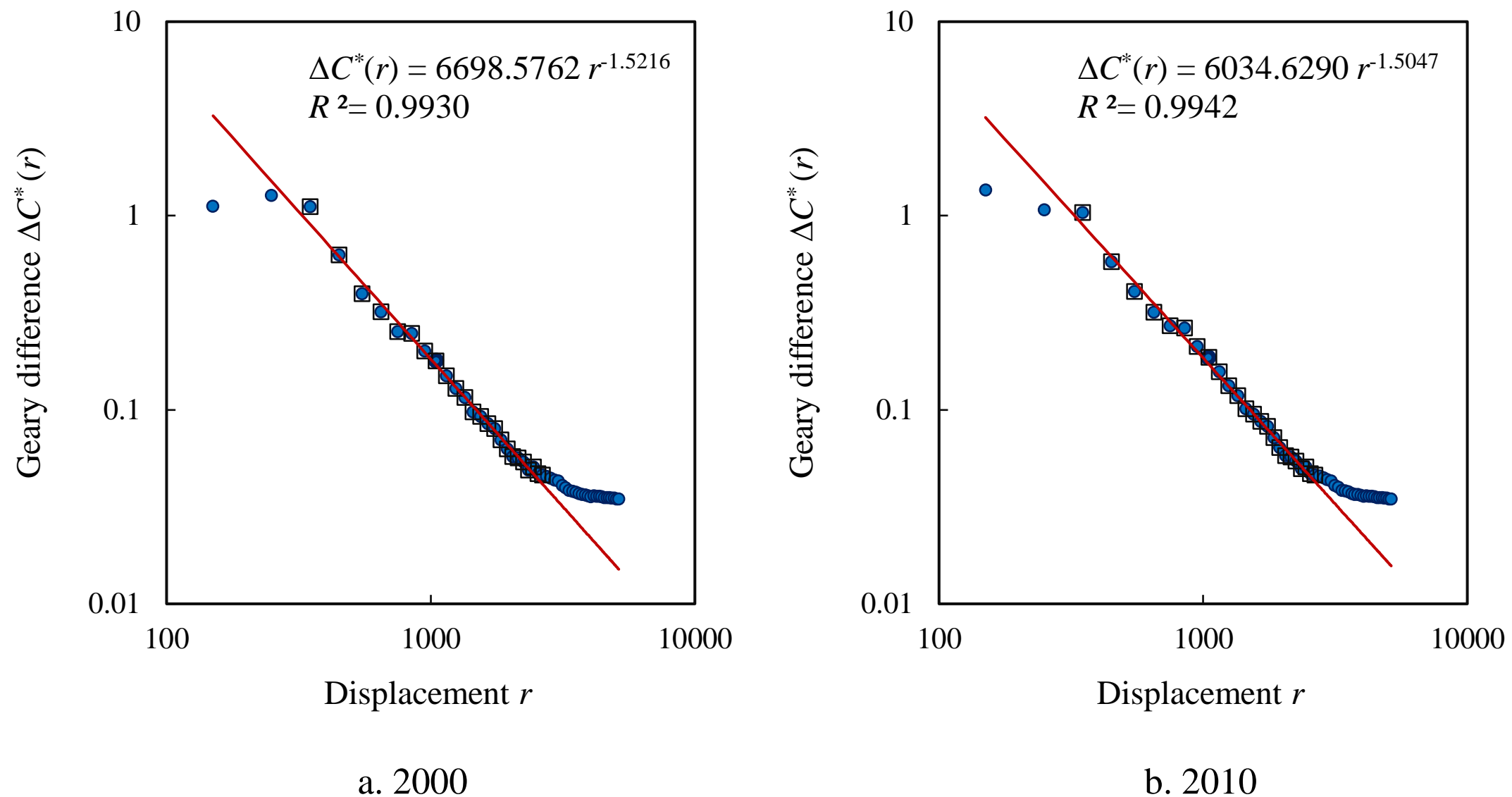

**Figure 6 The scaling relations for the difference of spatial autocorrelation functions based on Geary's coefficient**

**Note:** The solid dots represent the total number of spatial autocorrelation functions, and the hollow blocks represent the points within the scaling range. The scaling range corresponds to that in Figure 1.

Further, we can testify the relationship between the spatial correlation dimension and the spatial autocorrelation function based on Getis-Ord's index. This relationship is determined by both spatial weight matrix and size vector. But the size variable influence the proportionality coefficient instead of spatial correlation dimension. For the observational data in 2000, the model is as follows

$$\hat{f}(G(r)) = \frac{u^{\mathrm{T}}u}{\hat{N}(r)} = 3.6221r^{-1.3623}, \tag{51}$$

where $u^{\mathrm{T}}u$=0.0555. The goodness of fit is about $R^2$=0.9965, and the spatial correlation dimension is about $D_c$=1.3623. For the data in 2010, the model is as below

$$\hat{f}(G(r))=\frac{u^{\mathrm{T}}u}{\hat{N}(r)}=3.6625r^{-1.3623}, \tag{52}$$

where $u^{\mathrm{T}}u$=0.0561. The goodness of fit and the spatial correlation dimension are the same as those in 2000, and they are also the same as those based on Moran's function. The fractal relation is valid only within certain scaling range (Figure 7), which is consistent with the scaling range reflected by spatial correlation dimension (Figure 1).

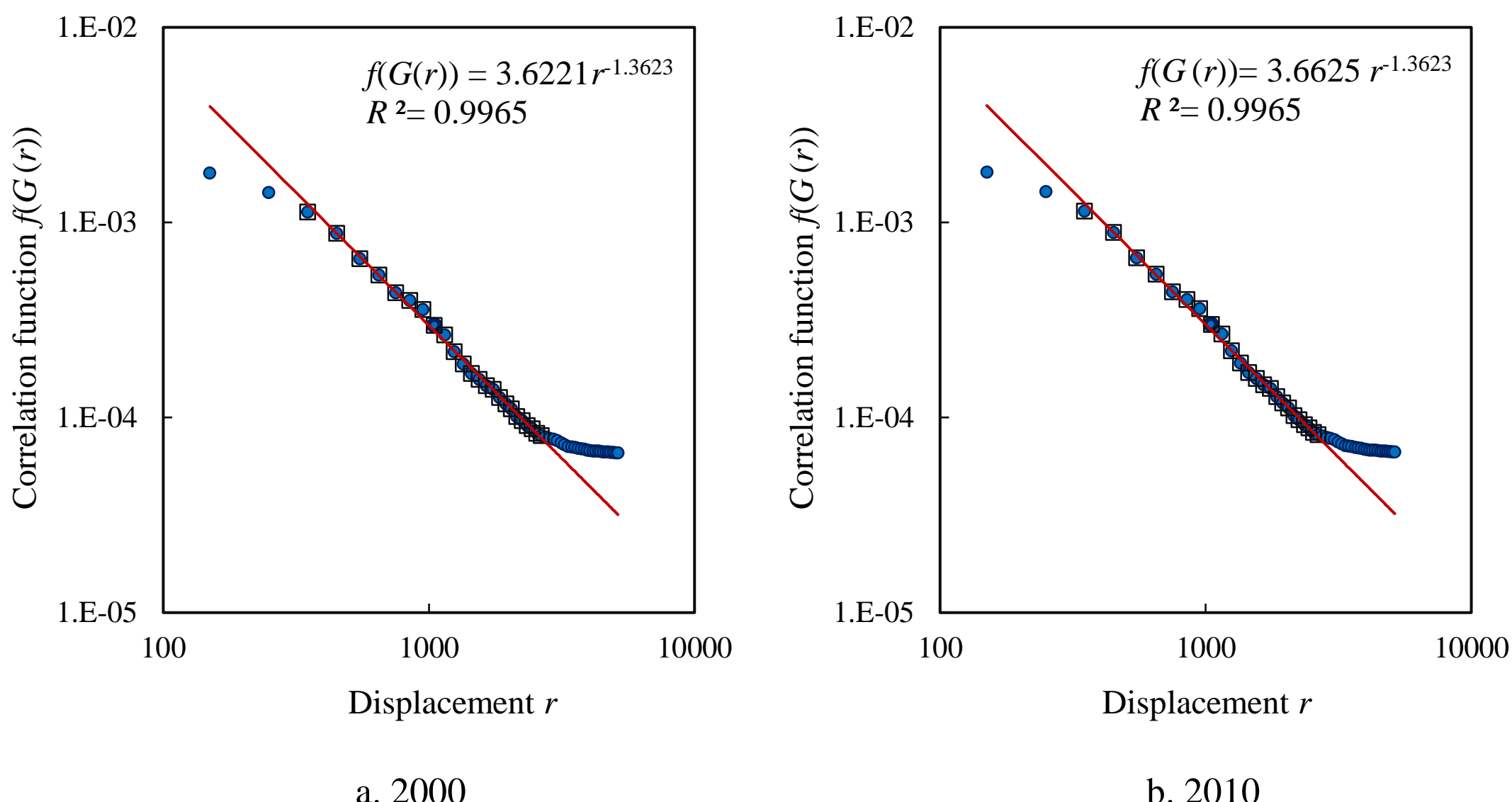

**Figure 7 The scaling relations for the generalized spatial correlation function based on Getis-Ord's index**

**Note:** The solid dots represent the total number of spatial autocorrelation functions, and the hollow blocks represent the points within the scaling range. The scaling range corresponds to that in Figure 1.

# 4 Discussion

The theoretical derivation and empirical analyses confirmed the mathematical and numerical relationships between the spatial correlation dimension and the generalized spatial autocorrelation functions. The ideas from spatial correlation are important in the research on both city fractals and fractal cities. As indicated above, one of fractal dimension definition is based on correlation functions. Spatial correlation can be divided into four types based on equation (10). If $r$ is a constant, we will have a correlation based on a fixed scale, which is used to define the common spatial autocorrelation coefficient; if $r$ depends on the size of geographical elements, we will have correlation based on characteristic scales; if $r$ is a variable but $i$ or $j$ is fixed to a certain element, we have a local scaling correlation, which can be used to define radial dimension of cities; if $r$ is a variable and $i$ and $j$ are not fixed to a certain element, we have a global scaling correlation, which can be used to define spatial correlation dimension derived above (Figure 8). The local correlation is termed one point correlation or central correlation, while the global correlation is termed point-point correlation or density-density correlation (Chen, 2013). The former reflects the 1-dimensional spatial correlation for isotropic development, while the latter reflect the 2-dimensional spatial correlation for anisotropic development. Spatial correlation is one of approaches to estimating fractal dimension of cities (Batty and Longley, 1994; Frankhauser, 1994; Frankhauser, 1998). A number of interesting studies have been made to calculate fractal dimension of urban form, and the method can be combined with dilation method (De Keersmaecker *et al*, 2003; Thomas *et al*, 2007; Thomas *et al*, 2008; Thomas *et al*, 2012). The spatial correlation analysis can be integrated into the percolation analysis to model the complex evolution of urban growth (Makse *et al*, 1995; Makse *et al*, 1998; Stanley *et al*, 1999). The above results form a bridge between spatial correlation of urban patterns and spatial autocorrelation of geographical processes by means of the concepts from fractals and scaling.

The spatial correlation dimension is one of basic parameter in the global fractal dimension set of multifractals. A multifractal system can be characterized with two sets of global and local parameters, which are connected with Legendre transform. The macro level of multifractals can be described with the generalized correlation dimensions and the corresponding mass exponents (Grassberger, 1983; Grassberger, 1985; Hentschel and Procaccia, 1983), and the micro level can be characterized with the local fractal dimensions and the corresponding singularity exponent (Frisch and Parisi, 1985; Halsey *et al*, 1986; Jensen *et al*, 1985). Multifractal geometry is one of powerful tools for geospatial analysis. The significant properties of geographical systems are dependence and heterogeneity (Anselin, 1996; Rey and Ye, 2010), and multifractal parameters are defined on the basis of entropy and correlation function (Chen, 2020; Feder, 1988; Grassberger, 1985; Liu and Liu, 1993; Stanley and Meakin, 1988; Wang and Li, 1996). The ideas from entropy can be used to deal with the spatial heterogeneity, while the notion from correlation function can be utilized to address the spatial dependence. Therefore, multifractal scaling not only represents a quantitative description method for a broad range of heterogeneous phenomena (Stanley and Meakin, 1988), but also an excellent approach to analyzing spatial dependence (Chen, 2013a). Cities proved to be complex spatial systems of the geographical world (Allen, 1997; Batty, 2005; Chen, 2008b; Portugali, 2011; Wilson, 2000). A system of cities proved to be a complex network with cascade and hierarchical structure (Batty and Longley, 1994; Frankhauser, 1998). Multifractal theory provides effective means for modeling complex network (Xue and Bogdan, 2017; Xue and Bogdan, 2019; Yang and Bogdan, 2020). The two central concepts in complexity science are emergence and dynamics (Batty, 2000). Multifractal geometry can be used to quantify emergence (Balaban *et al*, 2018), and spatial autocorrelation measures can be used to explore spatial dynamics (Rey and Ye, 2010). Multifractal modeling has been applied to urban and regional studies (Appleby, 1996; Ariza-Villaverde *et al*, 2013; Cavailhès *et al*, 2010; Chen, 2008b; Haag, 1994; Hu *et al*, 2012; Murcio *et al*, 2015; Pavón-Domínguez *et al*, 2018; Semecurbe *et al*, 2016). Among the generalized correlation dimension set, there are three basic parameters: capacity dimension, information dimension, and correlation dimension (Grassberger, 1983). The three parameters are suitable for describing the three important aspects of urban systems (Table 5). The inherent relation between spatial correlation dimension and spatial autocorrelation function opens up a new way of understanding complex systems of cities.

**Table 5 The measures and meanings of three basic fractal parameters in the generalized correlation dimension spectrum for urban systems**

| **Fractal dimension** | **Basis of definition** | **Measure of space** | **Probability value** | **Geographical meaning** |
|---|---|---|---|---|
| **Capacity dimension $D_0$** | Hartley entropy | Degree of space filling | Categorical variable ($P$=0, or $P$=1) | Is there a city in a place? |
| **Information dimension $D_1$** | Shannon entropy | Degree of spatial uniformity | Metric variable ($0 \leq P \leq 1$) | How many cities are there in a place? |
| **Correlation dimension $D_2$** | The second order Renyi entropy | Degree of spatial dependence | (1) Categorical variable ($P$=0, or $P$=1);<br>(2) Metric variable ($0 \leq P \leq 1$) | (1) If a city is found in one place, can another city be found within a given distance?<br>(2) If a city is found in a place, what is the probability of finding another city in a given distance? |

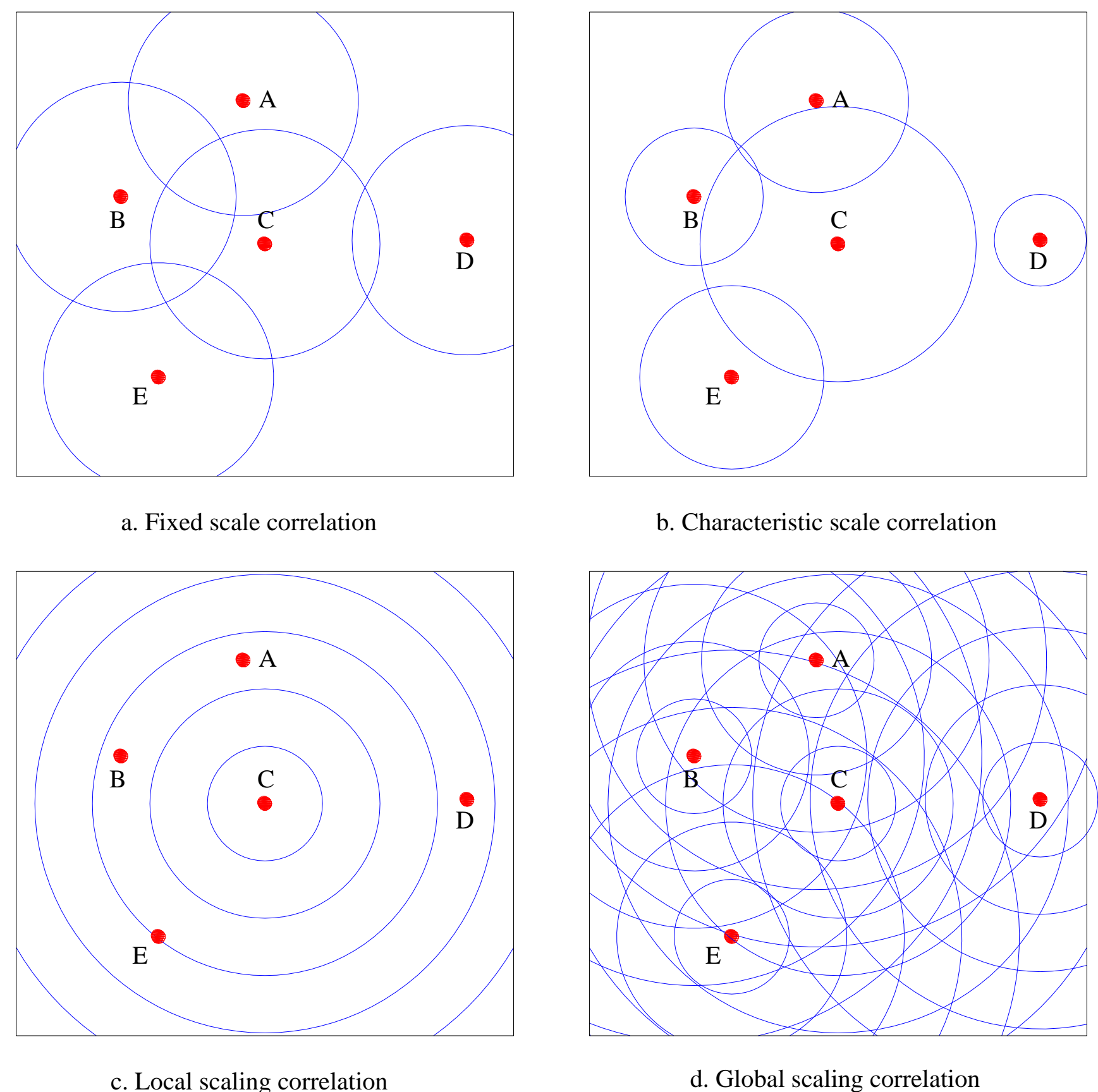

a. Fixed scale correlation

b. Characteristic scale correlation

c. Local scaling correlation

d. Global scaling correlation

**Figure 8 A sketch map of spatial correlation which fall in four types**

**Note:** The spatial correlation based on fixed scale can be used to calculate Moran's index, the one point correlation based on local scaling can be used to compute radial fractal dimension, and the point-point correlation based on global scaling can be used to calculate spatial correlation dimension and define spatial autocorrelation function.

The novelty of this paper lies in deriving the mathematical relationships between spatial autocorrelation functions and spatial correlation dimension. Where cities are concerned, the fractal dimension of spatial correlation depends heavily on the spatial distribution rather than size distribution of cities. The shortcoming of this work lies in two respects. First, the empirical analyses are based on 29 provincial capital cities rather than a system of cities based on certain size threshold. The system of provincial capital cities are in the administrative sense instead of pure geographical sense. This type of spatial sample can be used to produce example to illustrate a research method. If we perform a spatial analysis of Chinese cities for practical problems, we should extract a spatial sampling according to certain scale threshold. Second, the case study is only based on the observational data of Chinese cities. If we can obtain the spatial dataset of other countries, maybe we can make a comprehensive positive studies. Unfortunately, due to the limitation of observed data as well as the space of a paper, the work remains to be done in the future.

## 5 Conclusions

For the complex spatial systems, the spatial autocorrelation coefficients face a dilemma. If a spatial autocorrelation coefficient is valid, it indicates no other useful spatial information except for no autocorrelation. In contrast, if the autocorrelation coefficient suggests significant correlation, the value is not so valid. The property of spatial autocorrelation influences the accuracy of the calculation of spatial autocorrelation coefficient itself such as Moran's index. The problem comes from spatial scaling, which impacts on mean, and thus on calculation result. In this case, spatial autocorrelation coefficient should be replaced by spatial autocorrelation functions. One of simple and important approach to constructing spatial autocorrelation functions based on spatial autocorrelation coefficients is to make use of the relative step function based on variable distance threshold. Thus, we can derive the spatial correlation dimension from the spatial autocorrelation functions. The main conclusions of this study can be reached as follows. **First, the spatial correlation dimension can be calculated by means of the relationships between the standard spatial autocorrelation function and the generalized spatial autocorrelation function**. The spatial autocorrelation coefficients are not enough to reflect the complex dynamics process of geographical evolution. Spatial autocorrelation functions can be employed to characterize the spatio-temporal dynamics of geographical systems, but the measurement procedure and quantitative description are complicated. Using spatial correlation dimension, we can condense sets of spatial parameters into a simple number, and thus it is easy to make spatial analyses of geographical processes. **Second, the spatial correlation dimension depends on spatial contiguity matrix rather than the size measure of geographical element.** Changing size measure such as city population does not influence the relationships between spatial autocorrelation functions and spatial correlation dimension. However, changing distances between geographical elements in a region leads to different relationships between Moran's functions and yardstick length and thus results in different spatial correlation dimension values. This suggests that the common spatial correlation dimension depends on spatial distribution patterns instead of size distribution patterns. **Third, the scaling ranges of spatial correlation dimension reflect the geographical scope of spatial** **autocorrelation and interaction.** In theory, the spatial correlation dimension is absolute, but in practice, the spatial correlation dimension is a relative measure and is always valid within certain range of measurement scales. By means of log-log plots, the scaling range can be approximately identified visually. Using the residuals sequence plot of global double logarithmic linear regression model and the curve of goodness of fit of local double logarithmic linear regression model for spatial correlation dimension, we can identify the scaling range more objectively. The scaling range corresponds to the scope of positive autocorrelation reflected by the generalized spatial autocorrelation function based on Moran's index. This implies that the scaling range represents a quantitative criterion of spatial agglomeration of geographical distributions.

### Acknowledgements

This research was sponsored by the National Natural Science Foundation of China (Grant No. 41671167). The support is gratefully acknowledged.